\documentclass[cits]{PoS}

\usepackage{amsmath,amsfonts,amssymb,slashed}
\usepackage{booktabs}

\newcommand{\Order}{\mathcal{O}}

\newcommand{\mpp}{m_\text{p}}
\newcommand{\mn}{m_\text{n}}
\newcommand{\mpii}{M_{\pi^0}}
\newcommand{\mpi}{M_\pi}
\newcommand{\Fpi}{F_\pi}

\newcommand{\beq}{\begin{equation}}
\newcommand{\eeq}{\end{equation}}

\newcommand{\md}{m_\text{d}}

\newcommand{\eps}{\epsilon}

\newcommand{\pp}{\mathbf{p}}
\newcommand{\qq}{\mathbf{q}}
\newcommand{\kk}{\mathbf{k}}

\renewcommand{\Re}{\text{Re}\,}

\title{Isospin breaking in pion--deuteron scattering
and the pion--nucleon scattering lengths}

\ShortTitle{Isospin breaking in $\pi^-d$ scattering}

\author{\speaker{Martin Hoferichter}, $^{ab}$ Vadim Baru, $^{cd}$ Christoph Hanhart, $^{e}$ Bastian Kubis, $^a$ Andreas Nogga, $^e$ and Daniel R.~Phillips$^f$\\
\llap{$^{a}$} Helmholtz-Institut f\"{u}r Strahlen- und Kernphysik (Theorie), 
Bethe Center for Theoretical Physics, Universit\"at Bonn, D-53115 Bonn, 
Germany\\
\llap{$^{b}$} Albert Einstein Center for Fundamental Physics, Institute for Theoretical Physics,
	    Universit\"at Bern, CH-3012 Bern, Switzerland\\
       \llap{$^{c}$}
Institut f\"ur Theoretische Physik II, Ruhr-Universit\"at Bochum, 
D-44780 Bochum, Germany\\
\llap{$^{d}$} ITEP, 117218, B. Cheremushkinskaya 25, Moscow, Russia\\
\llap{$^{e}$} Forschungszentrum J\"ulich, Institut f\"ur Kernphysik, 
J\"ulich Center for Hadron Physics  and Institute for Advanced Simulation,  D-52425 J\"ulich, Germany\\
\llap{$^{f}$} Institute of Nuclear and Particle Physics and Department of Physics and Astronomy, Ohio University, Athens, OH 45701, USA\\
        E-mail: \email{hoferichter@itp.unibe.ch}}

\abstract{In recent years, high-accuracy data for pionic hydrogen and deuterium have become the primary source of information on the pion--nucleon scattering lengths. Matching the experimental precision requires, in particular, the study of isospin-breaking corrections both in pion--nucleon and pion--deuteron scattering. We review the mechanisms that lead to the cancellation of potentially enhanced virtual-photon corrections in the pion--deuteron system, and discuss the subtleties regarding the definition of the pion--nucleon scattering lengths in the presence of electromagnetic interactions by comparing to nucleon--nucleon scattering. Based on the $\pi^\pm p$ channels we find for the virtual-photon-subtracted scattering lengths in the isospin basis $a^{1/2}_{\slashed\gamma}=(170.5\pm 2.0)\cdot 10^{-3}\mpi^{-1}$ and 
$a^{3/2}_{\slashed\gamma}=(-86.5\pm 1.8)\cdot 10^{-3}\mpi^{-1}$.}

\FullConference{The 7th International Workshop on Chiral Dynamics,\\
		August 6 -10, 2012\\
		Jefferson Lab, Newport News, Virginia, USA}

\begin{document}

\section{Introduction}

Leading-order chiral perturbation theory (ChPT) predicts the low-energy theorem~\cite{Weinberg}
\beq
\label{LET}
a^-=\frac{\mpi}{8\pi(1+\mpi/\mpp)\Fpi^2}\approx 80\cdot 10^{-3}\mpi^{-1}
\eeq
for the isovector pion--nucleon ($\pi N$) scattering length $a^-$. While this prediction for $a^-$ is very stable against higher-order corrections, which only enter at $\Order(\mpi^3)$~\cite{BKM}, the chiral expansion for its iso\-scalar counterpart $a^+$ vanishes at leading order and involves large cancellations amongst the subleading terms. Moreover, given precise input for $a^-$, the Goldberger--Miyazawa--Oehme (GMO) sum rule~\cite{GMO} relates the $\pi N$ coupling constant to an integral over $\pi N$ cross sections.
Therefore, new, independent information on both the isovector and isoscalar scattering lengths 
becomes particularly interesting.

Such an approach is offered by hadronic atoms, more precisely high-precision measurements of the spectra of pionic hydrogen ($\pi H$) and deuterium ($\pi D$)~\cite{data_piH,data_piD}. These systems, composed of a $\pi^-$ and a proton/deuteron, are bound by electromagnetism, but strong interactions induce distortions of the pure QED spectrum. Accordingly, the level shift of the ground state is related to elastic $\pi^- p$ and $\pi^- d$ scattering, e.g.\ for the level shift $\eps_{1s}$ in $\pi H$ (with reduced mass $\mu_H$)
\beq
\label{Deser}
\eps_{1s}=-2\alpha^3 \mu_H^2a_{\pi^-p}\big( 1+2\alpha(1-\log \alpha)\mu_H a_{\pi^-p}+\cdots\big),
\eeq
while the width of the $\pi H$ ground state gives access to the charge-exchange reaction $\pi^-p\to\pi^0 n$~\cite{hadatoms}. This leads to the following system for $a^\pm$
\begin{align}
 a_{\pi^- p}&= \tilde a^+ + a^- + \Delta \tilde a_{\pi^- p},\qquad a_{\pi^- p\to\pi^0 n}=-\sqrt{2}\, a^- + \Delta  a_{\pi^- p\to \pi^0 n},\notag\\
 \Re a_{\pi^- d}&=2\frac{1+\mpi/\mpp}{1+\mpi/\md}(\tilde a^++\Delta \tilde a^+)+a_{\pi^- d}^{(3)},
\end{align}
where 
\beq
\tilde a^+ = a^+ + \frac{1}{4\pi(1+\mpi/\mpp)}
\Bigg\{\frac{4(\mpi^2-\mpii^2)}{\Fpi^2}c_1-2e^2f_1\Bigg\}
\eeq
includes isospin violation at leading order in ChPT~\cite{mrr}, $\Delta \tilde a_{\pi^- p}$, $\Delta  a_{\pi^- p\to \pi^0 n}$, and $\Delta \tilde a^+$ denote further isospin-breaking corrections~\cite{HKM,GR02}, and $a_{\pi^- d}^{(3)}$ incorporates three-body contributions in $\pi D$. The theoretical tool to evaluate these corrections is chiral effective field theory~\cite{Weinberg92,beane98,beane,recoil,disp+delta,Liebig,JOB,MSS}.
From the uncertainty estimate for the isospin-conserving three-body contributions of $1\cdot 10^{-3}\mpi^{-1}$~\cite{JOB}, compared to $\Re a_{\pi^- d}\sim -25\cdot 10^{-3}\mpi^{-1}$, it follows that only effects significantly below that threshold may be ignored.
In particular, at the level of accuracy required for the interpretation of the hadronic-atom data it becomes mandatory to investigate the role of isospin violation in $a_{\pi^- d}^{(3)}$~\cite{JOB}.

\section{Isospin breaking in threshold $\pi^- d$ scattering}

\begin{figure}
\begin{center}
\includegraphics[width=0.2\linewidth]{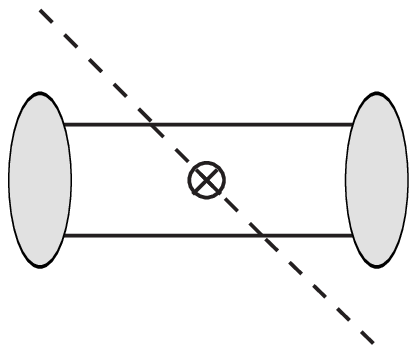}\qquad\qquad
\raisebox{6mm}{\includegraphics[width=0.2\linewidth]{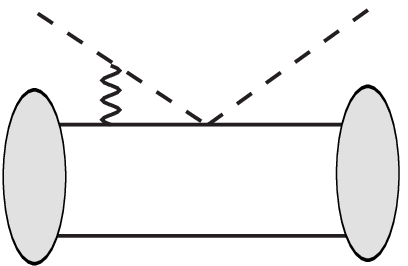}}
\end{center}
\caption{Mass-difference and virtual-photon effects in $\pi^- d$ scattering. Solid, dashed, and wiggly lines denote nucleons, pions, and photons, while the cross and the blobs refer to a mass-difference insertion and the deuteron wave function, respectively.}
\label{fig:IV}
\end{figure}

Isospin breaking is generated both by the difference between the light quark masses and virtual-photon effects, see Fig.~\ref{fig:IV} for representatives of each class. Mass-difference insertions are numerically relevant only for the leading, double-scattering diagram (first diagram in Fig.~\ref{fig:IV}), to which they contribute in the combination
\beq
\rho=2\mpi\Delta_{\rm N}-\Delta_{\pi},\qquad \Delta_{\rm N}=\mn-\mpp,\qquad \Delta_{\pi}=\mpi^2-\mpii^2.
\eeq
Since the leading-order pion mass difference is caused solely by electromagnetic effects, this implies that the quark mass difference only enters at subleading orders, and indeed $\Delta_\pi$ is responsible for the bulk of the total $2\%$ correction. In addition, isospin violation at the vertices, which may again be related to isospin breaking in the $\pi N$ scattering lengths, generates another $1\%$ effect.

The calculation of virtual-photon contributions to $a_{\pi^- d}^{(3)}$ proves particularly challenging due to the presence of various momentum scales: $p\sim \alpha\mpi$ (``hadronic-atom regime''), $p\sim \mpi$ (``chiral regime''), $p\sim\sqrt{\mpp\eps}$ (deuteron wave function), and $p\sim \sqrt{\mpi\eps}$ (three-body dynamics), with the deuteron binding energy $\eps$. While the hadronic-atom regime is already included in the calculation that leads to~\eqref{Deser}, the remaining scales might lead to an enhancement $\sim\sqrt{\mpi/\eps}$ of virtual-photon effects.
For isovector $\pi N$ interactions the pertinent integral takes the form
\beq
\label{isovector}
a_{T=1}\propto a^-\int d^3p\,d^3q\frac{\big(\Psi^\dagger(\pp-\qq)-\Psi^\dagger(\pp)\big)\Psi(\pp)}{\qq^2\big(\qq^2+2\mpi(\epsilon+\pp^2/\mpp)\big)},
\eeq
with the deuteron wave function $\Psi(\pp)$.
Indeed, the individual terms corresponding to $\Psi^\dagger(\pp-\qq)$ and $\Psi^\dagger(\pp)$ scale with $\sqrt{\mpi/\eps}$, but such contributions cancel in their difference~\cite{JOB}. The occurrence of this cancellation can be traced back directly to the Pauli principle, which
forces the intermediate $NN$ pair to be in a $P$-wave and thus leads to the relative sign in~\eqref{isovector}. 

In the isoscalar case intermediate-state $S$-wave $NN$ interactions are now permitted, in particular the deuteron pole that is already included in~\eqref{Deser} needs to be separated. Expressed in terms of overlap integrals between the deuteron and continuum wave functions $\Psi(\qq)$ and $\Psi_\pp^s(\qq)$, the contribution to the $\pi^- d$ scattering length then becomes 
\begin{align}
a_{T=0}&\propto a^+\int\frac{d^3k}{(2\pi)^3}\frac{1}{\kk^2}
\Bigg\{\frac{|F(\kk)|^2-1}{\kk^2/2\mpi-i\eta}+\frac{1}{2}
\int\frac{d^3p}{(2\pi)^6}
\frac{G_\pp^s(\kk)(G_\pp^s(\kk)+G^s_\pp(-\kk))}{\epsilon+\pp^2/\mpp+\kk^2/2\mpi-i\eta}\Bigg\},\notag\\
F(\kk)&=\int d^3q\Psi^\dagger(\qq)\Psi(\qq-\kk/2),\qquad
G_\pp^s(\kk)=\int d^3q\Psi^\dagger(\qq)\Psi_\pp^s(\qq-\kk/2),
\end{align}
which, by virtue of the normalization of $\Psi(\qq)$ and the orthogonality of bound-state and continuum wave functions for vanishing momentum transfer
\beq
|F(\kk)|^2-1=\Order(\kk^2),\qquad G_\pp^s(\kk)=\Order(\kk),
\eeq
proves the cancellation of the leading infrared enhanced contributions also for isoscalar $\pi N$ interactions. Explicit calculation shows that the infrared enhancement of momenta $\sim\sqrt{\mpp\eps}$ is too weak to become numerically relevant, so that in the end the only non-negligible correction due to virtual photons is generated by residual isovector terms dominated by momenta $\sim\mpi$~\cite{JOB}
\beq
a^{\rm EM}=(0.95\pm 0.01)\cdot 10^{-3}\mpi^{-1}.
\eeq
These findings
vindicate a posteriori the application of a chiral power counting and imply that the main impact of isospin violation for the extraction of the $\pi N$ scattering lengths is due to next-to-leading order isospin-breaking corrections, in particular the large shift $\Delta \tilde a^+=(-3.3\pm 0.3)\cdot 10^{-3}\mpi^{-1}$, in the $\pi N$ amplitude~\cite{HKM_CD09}. 

\begin{table}
\centering
\begin{tabular}{ccccc}
\toprule
isospin limit & channel & scattering length & channel & scattering length \\\midrule
$a^++a^-$ & $\pi^-p\rightarrow \pi^-p$ & $86.1\pm 1.8$ & $\pi^+n\rightarrow \pi^+n$ & $85.2\pm 1.8$ \\
$a^+-a^-$ & $\pi^+p\rightarrow \pi^+p$ & $-88.1\pm 1.8$ & 	$\pi^-n\rightarrow \pi^-n$  & $-89.0\pm 1.8$ \\
$-\sqrt{2}\,a^-$ & $\pi^-p\rightarrow \pi^0n$ & $-121.4\pm 1.6$ &  	$\pi^+n\rightarrow \pi^0 p$ & $-119.5\pm 1.6$\\
$a^+$ & $\pi^0p\rightarrow \pi^0p$ & $2.1\pm 3.1$ & 		$\pi^0n\rightarrow \pi^0 n$ &  $5.5\pm 3.1$\\
\bottomrule
\end{tabular}
\caption{$\pi N$ scattering lengths for the physical channels in units of $10^{-3}\mpi^{-1}$, Table taken from~\cite{JOB}.}
\label{table:physical_channels}
\end{table}

The final results for the combined analysis of $\pi H$ and $\pi D$ for the $\pi N$ scattering lengths are~\cite{JOB}
\beq
a^+=(7.6\pm 3.1)\cdot 10^{-3}\mpi^{-1}, \qquad \tilde a^+=(1.9\pm 0.8)\cdot 10^{-3}\mpi^{-1},\qquad
a^-=(86.1\pm 0.9)\cdot 10^{-3}\mpi^{-1},
\eeq
which, in combination with the isospin-breaking corrections from~\cite{HKM}, lead to the results for the physical channels given in Table~\ref{table:physical_channels}.

\section{Modified effective range expansion and subtraction of virtual-photon effects}

To illustrate the issues regarding the definition of a scattering length for charged particles in the presence of electromagnetic interactions we first consider the example of proton--proton scattering. First, the pure Coulomb phase shift $\sigma^C$ is removed from the total phase shift, so that the remainder $\delta^C_{pp}$, related to the strong amplitude $T_{pp}(k)$ by
\beq
k\big(\cot \delta^C_{pp}-i\big)=-\frac{4\pi}{m}\frac{e^{2i\sigma^C}}{T_{pp}(k)},\qquad k=|\kk|,
\eeq 
obeys the modified effective range expansion~\cite{Bethe}
\begin{align}
\label{mERE}
k\Big[C_\eta^2\big(\cot \delta^C_{pp}-i\big)+2\eta H(\eta)\Big]&=-\frac{1}{a_{pp}^C}+\frac{1}{2}r_0k^2+\Order(k^4),\\
C_\eta^2=\frac{2\pi\eta}{e^{2\pi\eta}-1},\qquad \eta&=\frac{\alpha m}{2k},\qquad
 H(\eta)=\psi(i\eta)+\frac{1}{2i\eta}-\log(i\eta),\qquad \psi(x)=\frac{\Gamma'(x)}{\Gamma(x)}.\notag
\end{align}
The removal of the residual Coulomb interactions to define a purely strong scattering length $a_{pp}$ is a scale-dependent procedure~\cite{JacksonBlatt,KongRavndal}
\beq
\frac{1}{a_{pp}}=\frac{1}{a_{pp}^C}+\alpha m\bigg[\log\frac{\mu\sqrt{\pi}}{\alpha m}+1-\frac{3}{2}\gamma_E\bigg],
\eeq
since the Coulomb-nuclear interference depends on the short-distance part of the nuclear force. 
Stated differently, for a consistent subtraction of virtual photons the electromagnetic coupling should be switched off also in the running of operators, which requires the knowledge of the underlying theory~\cite{Gegelia,GRS}. For $pp$ scattering such residual Coulomb effects induce a huge difference between $a_{pp}$ and $a_{pp}^C$~\cite{Bergervoet,Miller}
\beq
a_{pp}^C=(-7.8063\pm 0.0026)\,\text{fm},\qquad a_{pp}=(-17.3\pm 0.4)\,\text{fm}.
\eeq

The standard ChPT convention for the $\pi N$ scattering lengths~\cite{hadatoms}
\beq
e^{-2i\sigma^C}T_{\pi^-p}=\frac{\pi\alpha\mu_H a_{\pi^-p}}{k}-2\alpha\mu_H\big(a_{\pi^-p}\big)^2\log\frac{k}{\mu_H}+a_{\pi^-p}+\Order(k,\alpha^2),
\eeq
with the Coulomb pole $\propto 1/k$ and the term $\propto \log k/\mu_H$ first generated at one- and two-loop level,
can be matched to the modified effective range expansion~\eqref{mERE} by expanding first in $\alpha$, then in $k$
\beq
e^{-2i\sigma^C}T_{\pi^-p}=\frac{\pi\alpha\mu_Ha_{\pi^-p}^C}{k}-2\alpha\mu_H\big(a_{\pi^-p}^C\big)^2\bigg(\gamma_E+\log\frac{k}{\alpha\mu_H}\bigg)+a_{\pi^-p}^C+\Order(k,\alpha^2),
\eeq
and thus
\beq
a_{\pi^-p}=a_{\pi^-p}^C+2\alpha\mu_H\big(a_{\pi^-p}^C)^2(\log\alpha-\gamma_E)+\Order(\alpha^2).
\eeq
The correction term, involving the same $\log\alpha$ already present in~\eqref{Deser}, numerically evaluates to $-0.5\cdot 10^{-3}\mpi^{-1}$, which is still appreciably smaller than the uncertainty in $a_{\pi^- p}$ itself (see Table~\ref{table:physical_channels}).

The $\pi N$ scattering lengths are of particular interest to help determine subtraction constants that appear in the GMO sum rule or, more generally, a dispersive analysis of $\pi N$ scattering, see e.g.~\cite{JOB,DHKM}. In these applications, the derivation of the dispersion relations relies on the analyticity properties of the strong amplitude, so that the scattering lengths should be purified from any virtual-photons effects. Strictly speaking, the discussion of the $pp$ case shows that this cannot be achieved completely model-independently unless the underlying theory is known. In ChPT these subtleties appear in the regularization of UV divergent virtual-photon diagrams, where the separation between mass-difference and virtual-photon contributions to the low-energy constants requires the choice of a scale. However, taking as an example the combination $a_{\pi^-p}-a_{\pi^+p}$ needed for the GMO sum rule, we find for the virtual-photon corrections 
\beq
a_{\pi^-p}^\gamma-a_{\pi^+p}^\gamma=(2.1\pm 1.8)\cdot 10^{-3}\mpi^{-1},
\eeq
so that the scale dependence of the virtual-photon-subtracted scattering lengths
\beq
\label{piN_GMO}
a_{\pi^-p}^{\slashed\gamma}-a_{\pi^+p}^{\slashed\gamma}=(171.3\pm 2.4)\cdot 10^{-3}\mpi^{-1}
\eeq
should be entirely negligible. These effects are so much smaller than in $pp$ scattering since $\pi N$ scattering is perturbative, whereas the fine tuning in the nucleon--nucleon potential enhances any residual virtual-photon contributions.

Using~\eqref{piN_GMO} as input for the GMO sum rule we find for the $\pi N$ coupling constant 
$g_c^2/4\pi=13.7\pm 0.2$~\cite{JOB}.
Finally, we give the virtual-photon-subtracted scattering lengths in the isospin basis as derived from elastic  $\pi^\pm p$ scattering
\beq
a^{1/2}_{\slashed\gamma}\dot=\frac{1}{2}\big(3a_{\pi^-p}^{\slashed\gamma}-a_{\pi^+p}^{\slashed\gamma}\big) =(170.5\pm 2.0)\cdot 10^{-3}\mpi^{-1},
\qquad
a^{3/2}_{\slashed\gamma}\dot=a_{\pi^+p}^{\slashed\gamma} =(-86.5\pm 1.8)\cdot 10^{-3}\mpi^{-1},
\eeq
which are needed as input for a dispersive analysis of $\pi N$ scattering~\cite{DHKM}.

\end{document}